\documentstyle[12pt,epsf]{article}
\begin{document}
\begin{center}
{\large\bf Parton Distribution of Proton in a Simple Statistical Model}

\vspace{1cm}

Yong-Jun Zhang\footnote{zyj@pubms.pku.edu.cn}

Department of Physics, Peking University, Beijing 100871, China and \\
Institute of High Energy Physics, CAS,  Beijing 100039, China

\bigskip
Bing-Song Zou\footnote{zoubs@mail.ihep.ac.cn}

CCAST (World Laboratory), P.O.~Box 8730, Beijing 100080 and \\
Institute of High Energy Physics, CAS,  P.~O.~Box 918(4), Beijing 100039, 
China

\bigskip
Li-Ming Yang

Department of Physics, Peking University, Beijing 100871, China
\date{\today}
\end{center}

\begin{abstract}
Taking proton as an ensemble of quark-gluon Fock states and using
the principle of detailed balance, we construct a simple
statistical model for parton distribution of proton. The recent
observed Bjorken-$x$ dependent light flavor sea quark asymmetry
$\bar{d}(x)-\bar{u}(x)$ can be well reproduced by Monte Carlo
simulation as a pure statistical effect.
\end{abstract}

\bigskip
{\quad\bf PACS: 12.40.Ee, 12.38.Lg, 14.20.Dh, 14.65.Bt}

\vspace{1cm}

Proton is the simplest system in which the three colors of QCD
neutralize into a colorless object, but its internal quark-gluon
structure is still not well understood. The complication comes from
the presence of sea quarks in the proton. In all global analyses of
parton distribution in nucleons before 1990, a symmetric
light-quark ($\bar u$, $\bar d$) sea was assumed, based on the
usual assumption that the sea of quark-antiquark pairs is produced
perturbatively from gluon splitting\cite{Garvey}. However, a
surprisingly large asymmetry between the $\bar u$ and $\bar d$
sea quark distributions in the proton has been observed in recent
deep inelastic scattering\cite{NMC,HERMES} and Drell-Yan
experiments\cite{NA51,E866,NuSea}.

There have been many theoretical
attempts\cite{Kumano,Thomas,Peng,Nikolaev,Alberg}
trying to find the origins for this asymmetry. It is
believed\cite{Garvey,NuSea} that the asymmetry cannot be produced
from perturbative QCD and mesonic degrees of freedom play an
important role for the effect.

In this paper we follow a new idea\cite{yongjunzhang} for the
origin of the light flavor sea quark asymmetry to reproduce the
recent observed $\bar{d}(x)-\bar{u}(x)$ distribution\cite{NuSea}
with a simple statistical model. The basic idea in
Ref.\cite{yongjunzhang} is rather simple:  while sea
quark-antiquark pairs are produced flavor blindly by gluon
splitting, $\bar u$ quarks have larger probability to annihilate
than $\bar d$ quarks due to the fact that there are more $u$ quarks
than $d$ quarks in the proton, which hence causes the asymmetry.
Taking proton as an ensemble of quark-gluon Fock 
states\cite{Brodsky,Hoyer} and using
the principle of detailed balance for transitions between various
Fock states through creation or annihilation of partons, the
probabilities $\rho_{i,j,k}$ of finding the quark-gluon Fock
states $|\{uud\}\{i,j,k\}\rangle$ have been
obtained and given in Table 2 of Ref.\cite{yongjunzhang}, with
${i,j,k}$ the number of $\bar{u}u$ pairs, the number of $\bar{d}d$
pairs, the number of gluons, respectively. With the density matrix
$\rho_{i,j,k}$ for the quark-gluon Fock states
$|\{uud\}\{i,j,k\}\rangle$, the sea-quark flavor asymmetry was
calculated as $\bar d-\bar u\approx 0.124$, which is in
surprisingly agreement with the experimental data $\bar d-\bar
u=0.118\pm 0.012$. Encouraged by this success, here we want to
extend the model to calculate the Bjorken-$x$ distribution of
partons to study the $x$-dependence of the flavor asymmetry in the
nucleon sea, $\bar{d}(x)-\bar{u}(x)$, which has recently been well
measured by the FNAL E866/NuSea Collaboration\cite{E866,NuSea}.

For a quark-gluon Fock state $|\{uud\}\{i,j,k\}\rangle$,
the total number of partons is $n=3+2i+2j+k$.
If the $n$ partons were free particles without mutual interactions,
their momentum distribution $d\rho^F_n(p_1,\cdots ,p_n)$ would
simply follow the $n$-body phase space, {\sl i.e.},
\begin{equation}
d\rho_n^F(p_1,\cdots,p_n)=d\Phi_n(P;p_1,\cdots,p_n)
=\delta^4(P-\sum^n_{i=1} p_i)\prod^n_{i=1}
\frac{d^3p_i}{(2\pi)^32E_i}
\end{equation}
with $P$ the 4-momentum of the proton, $p_1,\cdots,p_n$ the
4-momenta of $n$ partons. If we ignore the mass of partons, then
$E_i\equiv\sqrt{\vec{p}\ ^2+m^2_i}=|\vec{p_i}|\equiv P_i$ and we
have
\begin{equation}
d\rho_n^F(p_1,\cdots,p_n)
=\delta^4(P-\sum^n_{i=1} p_i)\prod^n_{i=1}
\frac{P_idP_id\Omega_i}{2(2\pi)^3}
=\delta^4(P-\sum^n_{i=1} p_i)\prod^n_{i=1}
\frac{E_idE_id\Omega_i}{2(2\pi)^3}.
\end{equation}
However, we know that partons are not free particles and are
confined in the proton. In potential picture, partons are almost
free only near the center of the proton and their momenta decrease
when moving away from the center. Partons with smaller momenta at
the center stay longer close to the center while partons with
larger momenta at the center stay shorter around the center. So
compared with total free particles, partons in the proton should
have larger probability for smaller momenta. In this paper we
assume the $n$-parton momentum distribution $d\rho_n(p_1,\cdots
,p_n)$ to be
\begin{eqnarray}
d\rho_n(p_1,\cdots,p_n)
&=&\frac{1}{\prod^n_{i=1}P_i}d\Phi_n(P;p_1,\cdots,p_n)
\nonumber \\
&=&\delta^4(P-\sum^n_{i=1} p_i)\prod^n_{i=1}
\frac{dP_id\Omega_i}{2(2\pi)^3}
=\delta^4(P-\sum^n_{i=1} p_i)\prod^n_{i=1}
\frac{dE_id\Omega_i}{2(2\pi)^3}.
\end{eqnarray}
This is equivalent to assuming equal probability for
any energy configuration $(E_1,\cdots,E_n)$ of $n$ partons in the
proton.

With the $n$-parton momentum distribution $d\rho_n(p_1,\cdots
,p_n)$ and taking the light front formula for the Bjorken-$x$
\begin{equation}
x=\frac{E_{\rm parton}-p_{z\ {\rm parton}}}{M_{\rm proton}},
\label{x}
\end{equation}
we can easily get the $x$-distribution of partons, $\rho_n(x)$, for
an $n$-parton Fock state of the proton by using a simple Monte Carlo
simulation with a Monte Carlo event generator program called
GENEV from the CERN computer program library (CERNLIB). The $\rho_n(x)$
is normalized as $\int^1_0dx\rho_n(x)=1$. 

For a $n$-parton Fock state $|\{uud\}\{i,j,k\}\rangle$, the total
number of partons is $n$ including $u, d, \bar{u}, \bar{d}$ and
$g$ with parton number of  $2+i$, $1+j$, $i$, $j$ and $k$, respectively. 
Then the $x$-distribution for each kind of partons, $u_{i,j,k}(x)$,
$d_{i,j,k}(x)$, $\bar{u}_{i,j,k}(x)$,
$\bar{d}_{i,j,k}(x)$ and $g_{i,j,k}(x)$, are
\begin{eqnarray}
u_{i,j,k}(x)&=&\rho_n(x)(2+i), \\
d_{i,j,k}(x) &=& \rho_n(x)(1+j), \\
\bar{u}_{i,j,k}(x) &=&\rho_n(x) i, \\
\bar{d}_{i,j,k}(x) &=& \rho_n(x)j, \\
g_{i,j,k}(x) &=& \rho_n(x)k. 
\end{eqnarray}

Summing up parton $x$-distribution of all possible Fock states
$|\{uud\}\{i,j,k\}\rangle$ according to their weights
($\rho_{i,j,k}$) in Table 2 of Ref.\cite{yongjunzhang} to get the parton
$x$-distribution of proton, we have
\begin{eqnarray}
u(x)=\sum_{i,j,k}\rho_{i,j,k}u_{i,j,k}(x),\\
d(x)=\sum_{i,j,k}\rho_{i,j,k}d_{i,j,k}(x),\\
\bar{u}(x)=\sum_{i,j,k}\rho_{i,j,k}\bar{u}_{i,j,k}(x),\\
\bar{d}(x)=\sum_{i,j,k}\rho_{i,j,k}\bar{d}_{i,j,k}(x),\\
g(x)=\sum_{i,j,k}\rho_{i,j,k}g_{i,j,k}(x),
\end{eqnarray}
which satisfy normalization condition
\begin{eqnarray}
\int_0^1 x\left[ u(x) + d(x) + \bar{u}(x)
  + \bar{d}(x) + g(x) \right] \, dx = 1.
\end{eqnarray}
Our $u(x)$ and $d(x)$ include both valence and intrinsic sea
quarks which are identical and not distinguishable in our
approach. The $x$-distribution of valance quarks ($u_v$, $d_v$)
can be easily obtained as
\begin{eqnarray}
u_v(x)&=&u(x)-\bar{u}(x), \\
d_v(x)&=&d(x)-\bar{d}(x).
\end{eqnarray}
In addition, we have
\begin{eqnarray}
\bar{n}&=&\int_0^1 \left[ u(x) + d(x) + \bar{u}(x)
  + \bar{d}(x) + g(x) \right] \, dx=5.57,\\
\bar{E}&=&\frac{M_{\rm Proton}}{\bar{n}}=\frac{0.938{\rm
GeV}}{5.57}=0.168 {\rm GeV},
\end{eqnarray}
where $\bar{n}$ is the average number of partons in proton and
$\bar{E}$ is the average energy of partons in proton.

\begin{figure}[htbp]
  \begin{center}
    \mbox{\epsfxsize=9.0cm\epsfysize=8.0cm\epsffile{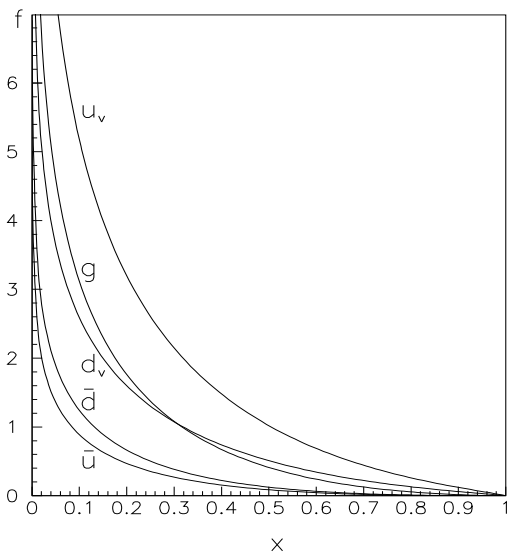}}
  \end{center}
\caption{The parton density $f$ $(f=u_v,\, d_v,\, 
\bar{u},\,
\bar{d},\, g)$ of our model at a scale $\mu_0\approx 0.168$ GeV.
The $f(x)$ in our model is simulated using Monte Carlo without any
parameter. All we need in our model are principle of detailed
balance and assumption of equal probability for every energy 
configuration of an $n$-parton Fock state.\label{density}}
\end{figure}

From above equations, we get the parton $x$-distribution densities 
$f(x,\mu_0^2)$ $(f=u,\, d,\, \bar{u},\, \bar{d},\,g)$ as shown in 
Fig.\ref{density} with the scale $\mu_0\approx\bar E=0.168$ GeV.

\begin{figure}[htbp]
  \begin{center}
\mbox{\epsfxsize=8.0cm\epsffile{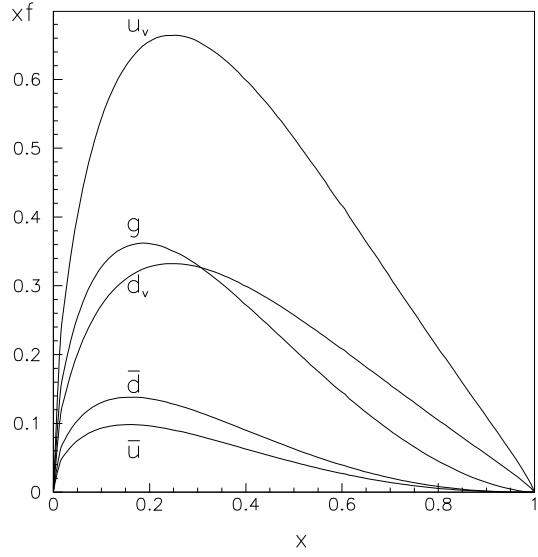}}
  \end{center}
\caption{The densities $xf$ $(f=u_v,\, d_v,\, 
\bar{u},\, \bar{d},\,
g)$ of our model at a scale $\mu_0\approx 0.168$ GeV. }
\label{momentum}
\end{figure}

\begin{figure}[htbp]
  \begin{center}
\mbox{\epsfxsize=11.5cm\epsffile{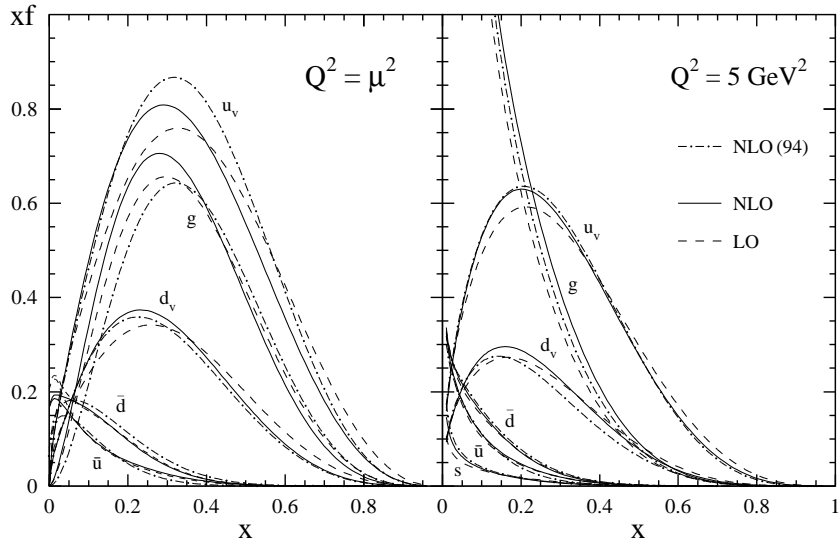}}
  \end{center}
\caption{For a comparison, the xf(x) of GRV\cite{GRV} at higher $Q^2$. 
Left: input densities $xf$ $(f=u_v,\, d_v,\, \bar{u},\, \bar{d},\, g)$
at $Q^2=\mu_{\rm LO}^2=0.26$ GeV$^2$ and $Q^2=\mu_{\rm
NLO}^2=0.40$ GeV$^2$, at which scales strange sea $s=\bar{s}$ vanishes;
Right: the evolved results at $Q^2=5$ GeV$^2$.}
\label{GRV}
\end{figure}

The corresponding momentum distribution, $xf(x,\mu_0^2)$, is shown 
in Fig.\ref{momentum}. As a comparison, the $xf(x,Q^2)$ distribution of 
GRV\cite{GRV} 
at some higher $Q^2$ is shown in Fig.\ref{GRV}.
It will be interesting to check whether 
with some QCD evolution equation our $xf(x,\mu_0^2)$ distribution could 
evolve into a momentum distribution $xf(x,Q^2)$ at higher $Q^2$ to be 
similar to the GRV's.

\begin{figure}[htbp]
  \begin{center}
    \mbox{\epsfxsize=10.cm\epsffile{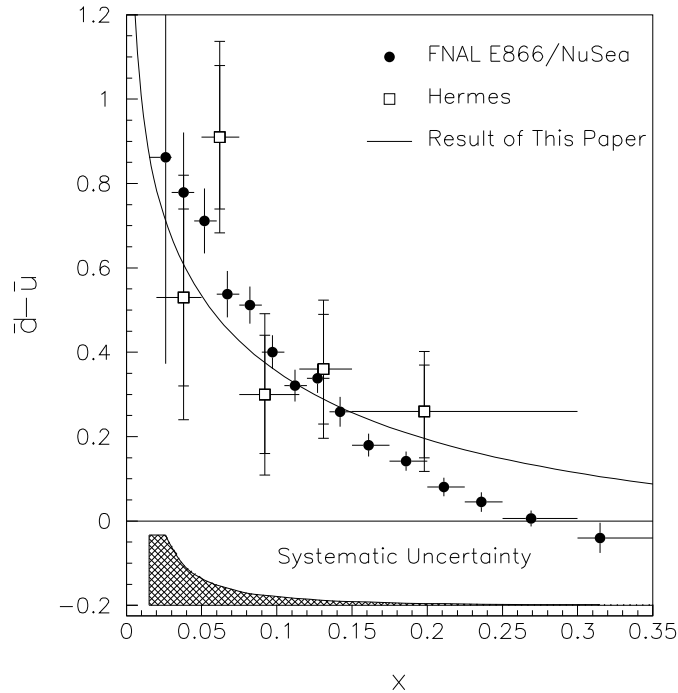}}
  \end{center}
\caption{Comparison of measured $\bar{d}(x)-\bar{u}(x)$ to prediction of
our model at scale $\mu_0\approx 0.168$GeV. The FNAL E866/NuSea results, 
scaled to fixed  $Q^2=54$GeV$^2$, are shown as the circles;  
HERMES results of $<Q^2>=2.3$GeV$^2$ are shown as squares. 
\label{asymmetry}}
\end{figure}

The quarks and gluons in the Fock states are the ``intrinsic" partons of 
the proton, multi-connected non-perturbatively to the 
valence quarks\cite{BM}. Such partons are different from ``extrinsic" 
partons generated from the QCD hard bremsstrahlung and gluon-splitting as 
part of the lepton scattering interaction. Partons measured at 
certain $Q^2$ by experiments include both ``intrinsic" and ``extrinsic" 
ones. Since ``extrinsic" partons are generated flavor blindly, 
the light flavor sea quark asymmetry is mainly due to ``intrinsic" partons 
and is practically $Q^2$ independent, although correlations between 
``intrinsic" and ``extrinsic" partons can cause some small $Q^2$ 
dependence for the asymmetry.  Experimental data at various $Q^2$ values
also show little 
$Q^2$ dependence\cite{Garvey,NuSea}. Hence we can compare our model
prediction of $\bar{d}(x)-\bar{u}(x)$ at the low scale $\mu_0$ 
directly with experimental data at higher scales, as shown in 
Fig.\ref{asymmetry}. 
Our prediction is in good agreement with recent experiment data of FNAL
E866/NuSea\cite{NuSea} at $Q^2=54$ GeV$^2$ and HERMES\cite{HERMES} at
$<Q^2>=2.3$ GeV$^2$.

In summary, following Ref.\cite{yongjunzhang} we take proton as an 
ensemble of quark-gluon Fock state,
{\sl i.e.}, $|p\rangle=\sum_{i,j,k}c_{i,j,k}|\{uud\}\{i,j,k\}\rangle$, 
with $\rho_{i,j,k}\equiv |c_{i,j,k}|^2$ determined by the principle of 
detailed balance. By further assuming equal probability 
for any energy configuration ($E_1,\cdots,E_n$) of $n$-parton Fock state 
in the proton, we get parton $x$-distribution functions at a scale of
$\mu_0\approx \bar E=0.168$ GeV with a Monte Carlo simulation of a simple 
statistical model. The corresponding light flavor sea quark 
asymmetry $\bar{d}(x)-\bar{u}(x)$ reproduces the recent experiment data 
quite well. This is a further support of the new origin of the light 
flavor sea quark asymmetry as a pure statistical effect due to the 
fact that there are more $u$-quarks than $d$-quarks in the proton.
\bigskip

{\bf Acknowledgment:} The work is partially supported by the CAS 
Knowledge Innovation Project (KJCX2-N11) and by National Natural 
Science Foundation of China.

\end{document}